# IS THE DENSITY DISTRIBUTION OF CLUSTERS NON-GAUSSIAN ?

Ts. Kolatt [1]

[1] *Harvard-Smithsonian Center for Astrophysics, Cambridge, Massachusetts, 02138 USA.*

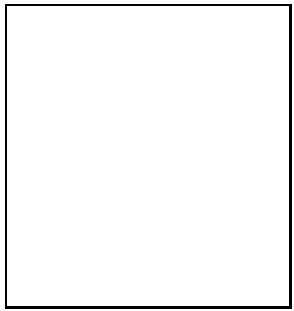

**Abstract**

The one-point probability distribution function ($PDF$) is computed for the 25 h$^{-1}$Mpc-smoothed density field of rich clusters of galaxies in the Abell/$ACO$ catalogs. The observed $PDF$ is compared to the $PDF$s drawn similarly from mock catalogs of clusters in cosmological simulations of Gaussian and several non-Gaussian initial conditions. Several statistics allow significant rejection of the non-Gaussian models tested here, and fail to reject the Gaussian model. A comparison with the predictions of second-order perturbation theory and a log-normal model for $CDM$ Gaussian initial conditions yield a linear biasing factor $b_c/b_o \simeq 4$ and $b_c/b_o \simeq 3.7$ for $R \geq 0$ clusters.

## 1 Introduction

There exist various cosmological scenarios that consider different physical processes that led to the initial mass-density fluctuations. All scenarios agree in that starting at a certain time gravitational instability took over and became the dominant process for the growth of density fluctuations. As for the initial conditions in different models, these can be categorized according to the Gaussian nature of their $PDF$. In most models that arise from the inflating universe paradigm [9] Gaussian density fluctuation fields originate naturally. Within the same framework of inflation some scenarios do lead to non-Gaussian density fields [15, 11]. In other models cosmic strings, monopoles and textures (see [5] for a review) all serve as high-density seeds for gravitational instability to start with, thus forming non-Gaussian fields. Some of these non-Gaussian models are motivated by observations [6, 16].

Assuming the initial fluctuations form a Gaussian random field, one may use either the gravitational potential field, the velocity field or the density field in order to reveal the Gaussian nature of them. The former two, however, are not yet well determined within a large enough volume to allow for conclusive results about their distribution function. The analysis of the spatial distribution of the density fluctuations is somewhat easier. By employing a biasing scheme we may use the object field as a probe for the distribution function. Since the density field may appear highly non-linear on small scales while we are looking for the (semi) linear phase of the distribution function, heavy smoothing is compulsory. This requirement, in conjunction with the requirement of having as many independent volumes as possible for the assessment of the PDF, lead to the natural choice of a big volume-limited sample of objects. The best candidates for this are the rich-cluster catalogs.

The direct calculation of a smoothed field PDF is preferential over the evaluation of n-order correlations (very noisy) or moments of distribution for discrete objects (subject to selection effects).

In the following we derive the PDF of the smoothed field of the rich-cluster catalogs (§2), compare with mock catalogs drawn from simulations with well known PDFs (§3), and demonstrate the rejection power of this test for some non-Gaussian PDFs (§4). We finally use the comparison between the first few moments as observed and as predicted-by-theory in order to obtain the linear biasing factor for the rich clusters.

## 2 Data and measuring method

Three combinations of two data sets were used in order to estimate the PDF of the rich clusters; The north Abell (NORTH) catalog [1] with richness class $R > 0$, the ACO [2] catalog, and the unification of the two sets including many more measured redshifts ($R > 0$ and $R \geq 0$).

Distances to the clusters, $D_c$, included in the NORTH and ACO sets are inferred by the procedure described in Scaramella et al. [21]. For the unified set of the above two (ALL), we use the compilation done by Peacock & West [18] and follow their procedure of distance estimation. In the present analysis we confine ourselves to include clusters within $D_c < 450$ h$^{-1}$Mpc . For the angular selection, $\phi(\mathbf{r})$, we use the formula of Bahcall & Soneira [3] for the NORTH sample, and Scaramella et al. [21] for atmospheric obscuration. The robustness of the results as a function of the different $D_c$ and $\phi(\mathbf{r})$ is discussed in detail elsewhere [13, 14]. The similar results for the PDF for each data set when analyzed separately (fig. 1) is due to this robustness. For the sake of the statistical noise reduction, we prefer to use the ALL data set.

We evaluate the 25 h$^{-1}$Mpc Gaussian smoothed PDF on grid points of 30 h$^{-1}$Mpc linear dimension. A key point is that the correction for selection is done for a smoothed field, any other choice doesn't allow appropriate weighting of low density cites. We weigh by setting $N_g = 50$ as the mass-unit measure of the clusters, where $N_g$ is the number count of galaxies within the Abell radius. Alternative weighting is discussed in [13]. In order to account for missing volume near the survey boundaries, the effective smoothing volume about each grid point, $V_{eff}$, is computed. The density assigned to a grid point is the weighted count of clusters times $V_{eff}^{-1}$. The PDF and its first moments are obtained from the distribution of smoothed density values at the grid points. In order to overcome the systematic errors and to reduce the random ones, we consider only regions in space where the selection function is not too low ($\phi \geq 0.4$) and points for which the effective window volume within the catalog boundaries is relatively large ($V_{eff} \geq 0.8V$).

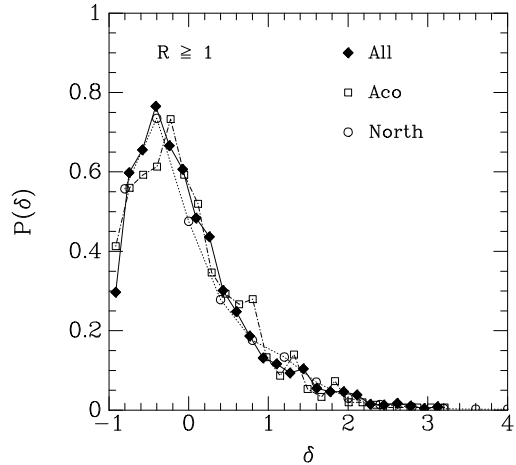

Figure 1: *PDF*s of the 25 h$^{-1}$Mpc smoothed three data sets ($D_c \leq 450$ h$^{-1}$Mpc , $R \geq 1$)

## 3 Comparison with simulations

### 3.1 Gaussian simulations

In order to quantify the rejection level of the hypothesis that the smoothed density distribution of the rich clusters of galaxies reconciles with Gaussian initial fluctuations, we create artificial catalogs out of N-body simulations. We use Zel'dovich mapping [25] for the simulations as an approximation for the mildly non-linear regime (for the quality of this *cf.* [12]). We take the power spectrum to be the 3 h$^{-1}$Mpc Gaussian smoothed *CDM* ($\Omega = 1, h_0 = 0.5, n = 1$) model [8] within a box of 900 h$^{-1}$Mpc side and $128^3$ particles and grid points. The identification of the considered cell as a cluster or a center for a super cluster, is done by applying the joint condition of both exceeding a threshold density and being a local peak of the density field.

We weigh each peak for which $\delta \geq \delta_{th}$ by $\delta - \delta_{th} + 1$. This allows to take the richness degree of freedom into account or alternatively to account for dense regions with more than one cluster. The specific locations of the clusters are then determined randomly within the boundaries of the considered cell. The observed number density is obtained by adjusting $\delta_{th}$.

In the upper left panel of figure 2 we notice the similarity between the *PDF* of the *ALL* clusters and the *PDF* from the simulation of the Gaussian field. Both *PDF*s are not Gaussian and several reasons account for that: **1.** Small non-linearities tend to change the original Gaussian density field. **2.** The biases due to the analysis contribute to mildly non-Gaussian effects. **3.** The identification of clusters is a process that may take us away from the original Gaussian nature of the density field [4].

The consistency of the *ALL* catalog distribution with the simulation's catalog, should, however, convince us that there is no contradiction between the two fields after they experience the measurement machinery. In order to make sure that such an agreement is not achieved by the combination of a high degree of statistical noise and heavy smoothing we should find at least one density field for which its derived distribution function differs considerably from the processed Gaussian field.

### 3.2 Non-Gaussian simulations

We divide the non-Gaussian fields into two categories: Local non-Gaussian where a local, continuous, non-linear transformation (mapping) is applied to a Gaussian field, and Global non-Gaussian where the construction of the field is by a process that induces strong correlations between the *phases* in the $k$-space distribution.

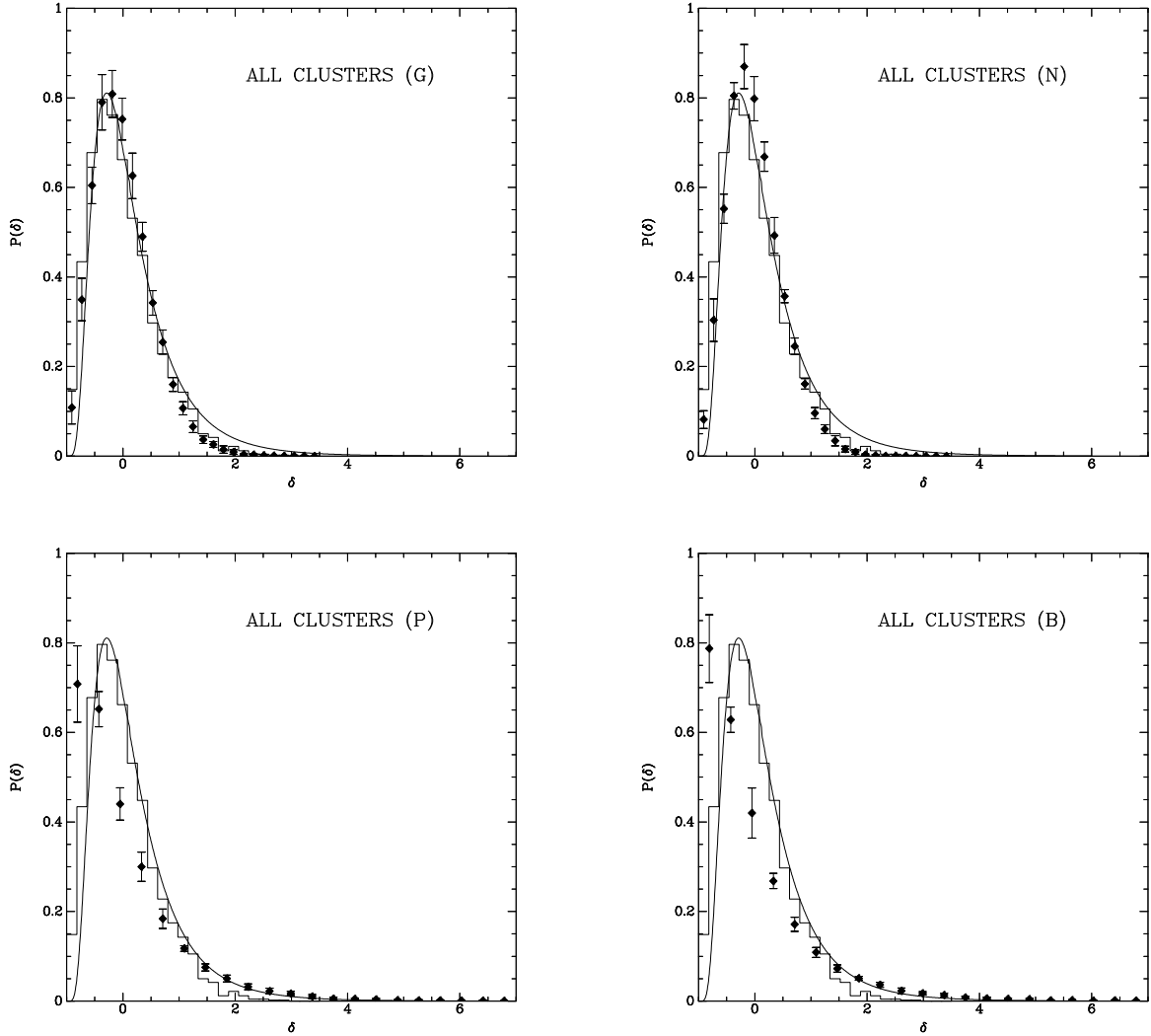

Figure 2: *PDF* drawn from 10, 25 h$^{-1}$Mpc smoothed artificial catalogs (diamonds, 1$\sigma$ errors) Vs. real data (*ALL*, solid histogram) and log-normal model (solid line). all for $R \geq 0$. **a.** Non-distorted Gaussian (G) **b.** Negatively skewed (N) **c.** Positively skewed (P) **d.** Broaden (B) (distortion parameter for (b-d), $a = 0.8$).

*3.2.1 Local non-Gaussian fields.* We follow Weinberg & Cole [24] and parameterize the local distortion from Gaussianity by one parameter, $a$. The transformations are applied to the field $\delta(\vec{r})$ with a Gaussian *PDF* of $\sigma^2$ variance. We force the distorted distribution to preserve the first and second moments of the Gaussian distribution. Three distortion schemes are given by considering the following transformations: positive, negative and broad.

$$Dis(\delta) = \begin{cases} \delta(\vec{r}) & (G) \\ (\exp(a\delta/\sigma) - \exp(a^2/2)) \times \sigma \left(\exp(2a^2) - \exp(a^2)\right)^{-1/2} & (P) \\ -(\exp(-a\delta/\sigma) - \exp(a^2/2)) \times \sigma \left(\exp(2a^2) - \exp(a^2)\right)^{-1/2} & (N) \\ \delta \left(\exp(a\delta/\sigma) + \exp(-a\delta/\sigma)\right) \times \sigma \left(2[1 + (1 + 4a^2)\exp(2a^2)]\right)^{-1/2} & (B) \end{cases} \quad (1)$$

A value of $a = 0.8$ is adopted since it yields similar distribution to that of the global texture model [17] in the (P) case.

In figure 2 the *PDF*s of the Gaussian and distorted fields are shown against the *PDF* of the (*ALL*) rich clusters ($R \geq 0$). In each plot the continuous line is the one parameter log-normal fit.

*3.2.2 Global non-Gaussian fields.* In the global non-Gaussian density field we use a "toy" explosion model [23]. Spheres of equal radius are embedded at random. Clusters are positioned at each of the two intersection points of every three spherical shells. The radius of the spheres and their filling factor are adjusted to obtain the desired number density of clusters, and several of their other observed statistical properties such as the auto-correlation function. Figure 3 shows the *PDF* obtained for the bubbles simulations (BUB) with bubbles of 50 h$^{-1}$Mpc radius and a filling factor of 1.5 .

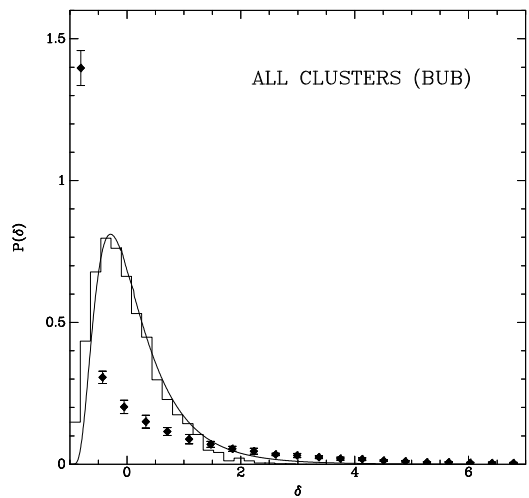

Figure 3: *PDF* from 30, 25 h$^{-1}$Mpc smoothed artificial bubbles catalogs (diamonds) Vs. *ALL* ($R \geq 0$) *PDF*. Marks as in figure 2.

## 4 Comparison with theoretical approximations

In this attempt, the 25 h$^{-1}$Mpc smoothing brings us to the slightly non-linear regime for which the second order perturbation theory is designed ([20] §18). In this framework the expression for the skewness, $S$, is obtained given a power spectrum, $P(k)$, and an arbitrary filter, $W_k$, [10]. For the standard *CDM* and the relevant current filter, numerical integration of the expression yields $S \approx 3.07\sigma$ (here $\sigma^2 \equiv \mu_2$).

A log-normal distribution may serve as an alternative approximation to the mildly non-linear evolution of the primary Gaussian distribution [7]. In this approximation we use the first (average, $\mu$) moment and the second (variance, $\sigma^2$) moment of the *Gaussian* distribution to express the entire moments of the *log-normal* distribution. In the current case where the first two moments should be preserved, the log-normal distribution becomes the mapping: $Dis(\delta) = \exp(\delta) - \exp(\sigma/2)$. Predictions for the skewness, by the two theoretical models, can be compared to the skewness as measured for the smoothed cluster density field. However the comparison may suffer from the following drawbacks: **1.** The approximations were made for the continuous field, while the data calculation is for a smoothed discrete field. **2.** The second order perturbation calculation underestimates the skewness. **3.** A time dependence may be required for the log-normal approximation [7].

Comparing an $S$ value as predicted by a model with the observed value according to $S^{model}(b_c\sigma) = S^{data}$ can then give a solution for $b_c$. In table 1 we quote the values of $b_c$ for standard *CDM* ($\sigma_8 = 1$, *i.e.* $b_o = 1$). From the second order perturbation theory we get $4.0 \pm 0.6$, and from the log-normal model $3.7 \pm 0.5$. The 1$\sigma$ errors are estimated from the simulations. These biasing factors are consistent with each other, and more important consistent with other calculations for rich cluster biasing factor ([19], $b_c/b_o = 4.5 \pm 0.6$).

**STATISTICS OF PDF OF DISTORTED FIELDS AND RICH CLUSTERS**

|  | $\sqrt{\mu_2}$ | $S_r$ | $K_r$ | $S$ | $K$ |
|---|---|---|---|---|---|
| **ALL CLUSTERS** | | | | | |
| ($R \geq 0$) | 0.58 | 1.47 | 4.34 | 1.00 | 1.14 |
| (G) | $0.52 \pm 0.02$ | $1.25 \pm 0.17$ | $4.36 \pm 0.04$ | $0.86 \pm 0.12$ | $1.07 \pm 0.45$ |
| **DISTORTED** | | | | | |
| (P) | $0.95 \pm 0.09$ | $1.28 \pm 0.11$ | $1.85 \pm 0.09$ | $2.43 \pm 0.70$ | $11.0 \pm 7.8$ |
| (N) | $0.48 \pm 0.02$ | $1.09 \pm 0.13$ | $4.40 \pm 0.03$ | $0.71 \pm 0.09$ | $0.50 \pm 0.21$ |
| (B) | $1.05 \pm 0.09$ | $1.39 \pm 0.07$ | $1.78 \pm 0.10$ | $2.78 \pm 0.67$ | $13.2 \pm 7.1$ |
| (BUB) | $1.61 \pm 0.12$ | $1.59 \pm 0.08$ | $1.67 \pm 0.11$ | $2.83 \pm 0.52$ | $10.96 \pm 4.75$ |
| CDM ($2^{nd}$ ORDER) | | | | $0.25$ ($b_c = 4.0 \pm 0.6$) | |
| LOG-NORMAL | | | | $0.27$ ($b_c = 3.7 \pm 0.5$) | |

Table 1: Statistics (see text) of the PDF for the real ($R \geq 0$) ALL data and artificial catalogs ($1\sigma$ errors).

**KOLMOGOROV - SMIRNOV REJECTION LEVEL (D)**

|  | ALL | ACO |
|---|---|---|
| (G) | NR($0.07 \pm 0.04$) | NR($0.07 \pm 0.04$) |
| (P) | $0.992(0.54 \pm 0.06)$ | $0.979(0.47 \pm 0.08)$ |
| (N) | NR($0.18 \pm 0.08$) | $0.956(0.34 \pm 0.06)$ |
| (B) | $0.999(0.58 \pm 0.03)$ | $0.997(0.51 \pm 0.02)$ |
| (BUB) | $0.999(0.56 \pm 0.00)$ | $0.999(0.63 \pm 0.01)$ |

Table 2: KS test rejection level ($D$) for ALL sample versus artificial catalogs ($R \geq 0$). NR≡Not-Rejected.

## 5 Statistical tests

One way of rejecting the hypothesis that two PDFs are consistent with each other is by comparing moments of the two distributions, rational functions of moments, or other statistics such as the "reduced skewness" $< \delta|\delta| >$. In Table 1 we show five statistics as applied to the $R \geq 0$ clusters. The statistics are: **(a)** standard deviation: $\sqrt{<\delta^2>} = \sqrt{\mu_2}$, **(b)** reduced skewness, defined as $S_r \equiv <\delta|\delta|>/\sqrt{\mu_2}$, **(c)** reduced kurtosis: $K_r \equiv (<|\delta|> - \frac{2}{\pi})/\sqrt{\mu_2}$, **(d)** skewness: $S \equiv [(\mu_3)^2/(\mu_2)^3]^{1/2}$, and **(e)** kurtosis: $K \equiv \mu_4/\mu_2^2$. The (P) model is formally ruled out by the $K_r$ statistic at the $\sim 27\sigma$ level. Similar results hold for the (B) model ($\sim 25\sigma$) and the (BUB) model ($\sim 24\sigma$). The (N) model can be rejected only at the $\sim 3\sigma$ level by $S_r$ and $S$ statistics. As for the (G) model, leaving aside the issue of $\sqrt{\mu_2}$, all the statistics yield a $\sim 1.5\sigma$ agreement between the data and the model. The $3\sigma$ deviation of $\sqrt{\mu_2}$ may be interpreted as the inadequacy of standard CDM power spectrum, or a biasing scheme somewhat different from the chosen one. The quoted numbers for the observational data can now serve as references for any theoretical model which predicts a certain pattern of a density field.

In order to allow consideration of the entire PDF in the comparison, we use the Kolmogorov-Smirnov (KS) test [22]. Table 2 sets the rejection levels drawn from the KS tests for the PDF of the rich clusters ($R \geq 0$), and those for the ensembles of artificial catalogs. Both PDFs are treated as experimental data. According to this test one can rule out consistency between the data and all the models but the (G) and the (N). In the case of the ACO catalog even the (N) model fails and the (G) model stays alone (NR = Not Rejected) to be consistent with the rich clusters distribution.

# 6 Conclusions

The smoothed density field of the Abell, ACO cluster catalogs provide a suitable probe for the mildly non-linear range of the evolution of density fluctuations, and may serve as a discriminatory tool for various initial distributions of matter fluctuations. In this contribution we confronted the smoothed PDF of the rich clusters with several non-Gaussian model PDFs and demonstrated its rejection power. No non-Gaussian model among those tested here (3 local non-Gaussian and 1 global) passed the test of being consistent with the data. We were unable to reject the Gaussian PDF model by any of the statistics we used here.

Our test is not a good discriminatory tool between different power spectra. In a biasing scheme it is subject to changes via multiplication by $b_c$. The power spectrum can however, be uniquely related to the other moments in the Gaussian case. Under the assumption of Gaussian field, the biasing factor can be found almost independently of the value of $\Omega$, by using second order perturbation theory ($b_c = 4.0 \pm 0.6$) or a log normal approximation ($3.7 \pm 0.5$).

This work was supported in part by the US National Science Foundation (PHY 91-06678).